# GENERALIZED ESTIMATES FOR THERMAL EXPANSION OF OXIDE SCALE IN THE RANGE FROM 0 ºC TO 1300 ºC WITH ACCOUNT FOR MOVABILITY OF PHASE TRANSITIONS IN ITS COMPONENTS

*Emmanuil Beygelzimer[1], Yan Beygelzimer[2]*

[1]*OMD-Engineering LLC, Dnipro, Ukraine*
[2]*Donetsk Institute for Physics and Engineering named after A.A. Galkin, National Academy of Sciences of Ukraine, Kyiv, Ukraine*

*Corresponding author: emmanuilomd@gmail.com, tel.: +380 (50) 368-63-42, 49000, Volodymyr Monomakh Street, 6, off. 303, Dnipro, Ukraine.

**ABSTRACT**

The thermophysical properties of oxide scale, in the general case, are affected by the variation of the temperature of phase transitions (either magnetic or polymorphic) in its components due to impurities, lattice defects, grain sizes, etc. In this case, since the phase transition is usually accompanied by a sharp change in properties, even a small shift of such a critical temperature can lead to large changes in properties in its vicinity. In order to account for this effect, data known from various sources on the true coefficient of linear thermal expansion (CLTE) of wüstite ($Fe_{1-x}O$), magnetite ($Fe_3O_4$), hematite ($Fe_2O_3$), and metallic iron are generalized by approximating functions that include critical temperatures as variable parameters. It is shown that the true CLTE of $Fe_3O_4$ at the same rated temperature can differ by up to 30% depending on the position of the Curie point within the limits of its possible "movability". The proposed methods allow to take into account the critical temperatures as adaptation parameters in engineering calculations of thermal expansion of oxide scale. Generalized formulas for each scale component are also given in a particular form for fixed (basic) values of critical temperatures. The dependence of the true CLTE of oxide scale as a whole ($α_{sc}$) on the volume fraction of each component is proposed. It is shown by model computations that there are temperatures at which $α_{sc}$ is almost independent of the scale composition ($α_{sc} ≈ 12·10^{-6}$ K$^{-1}$ at 250 °C and $α_{sc} ≈ 15·10^{-6}$ K$^{-1}$ at 650 °C), as well as areas of instability (400-600 °C and above 750 °C), where $α_{sc}$ depends significantly on the percentage of components. The results of the work are recommended to be used when mathematical modeling of production and processing of steel products in the presence of oxide scale on their surface.

## INTRODUCTION

In all high-temperature processes of production and use of steel products, in the presence of oxygen, oxide scale is formed, that, in general, consists of wüstite FeO (more precisely, $Fe_{1-x}O$, where $x ≈ 0.05...0.15$), magnetite $Fe_3O_4$, hematite $Fe_2O_3$, metallic iron, oxides of alloying elements and gas voids [1-3]. In many cases this scale has a decisive influence both on the process itself and on the quality of the product. For example, in hot rolling oxide scale affects the friction in the deformation zone, tool wear and the behavior of the metal at the roll exit [4-10]; scale on the rod determines the preparation of its surface before drawing [11-13]; scale significantly affects the intensity of heat transfer in the CCM secondary cooling zone, during thermomechanical control process (TMCP) and quenching [14-19]; oxide scale protects structural alloys against corrosion during liquid-metal cooling [20-21].

Therefore, recently the number of studies related both to the mathematical description of the behavior of oxide scale itself [7; 22-26], and with the modeling of technological processes, taking into account its presence on the surface of products [27-32], has increased sharply.

However, as a rule, the thermophysical properties of oxide scale are taken into account by some constants or estimated by particular data obtained under specific conditions, that significantly limits the predictive power of such models.

One reason for this situation is that information about the properties of the components of oxide scale are scattered in different references, are often fragmentary and in some cases do not agree with each other. For example, the true linear coefficient of thermal expansion of wüstite at 650 °C according to [33] and [34] differ by a factor of 2.5 (respectively $3.8·10^{-5}$ and $1.5·10^{-5}$ K$^{-1}$). Therefore, in engineering calculations, it is advisable to rely on property values obtained not in a specific experiment, but by generalizing the results of various studies. Comprehensive work on the systematization and generalization of the thermophysical properties of various substances was carried out by a team headed by Prof. Y. S. Touloukian in the 1960s and 70s [35]. However, since then, new experimental data on iron oxides have been accumulated and their critical states have been clarified (for example, the magnetic transition of $Fe_2O_3$ at temperatures around 1050 K (777 ºC), which had been suggested earlier on the basis of [37], was not confirmed [36]). In addition, large differences between the values of properties in different studies can be caused not only by peculiarities of experiments and inevitable heterogeneity of properties of different samples, but also by inconstancy ("movability") of critical temperatures, in the vicinity of which, as a rule, there are sharp changes in material properties. For example, the temperature of the magnetic transition of a crystalline substance may vary due to impurities, lattice defects, particle sizes, etc. [38]. The temperature of polymorphic transformation can vary even more widely, since it also depends on the rate of cooling (heating) [39]. Data on the critical temperatures, at which the peculiarities of changes in the properties of the main components of oxide scale, are summarized in **Table 1**. Given that even a small shift of the critical temperature can cause large changes in properties, it is advisable to include the critical temperatures in the approximating functions in the form of formal parameters. This will make it possible to use them as adaptation coefficients in mathematical modeling of technological processes and thereby take into account the influence of the movability of critical temperatures.

One of the main thermophysical characteristics of a solid matter is the true coefficient of linear thermal expansion (CLTE) which is understood as:

$$\alpha(T) = \frac{1}{L} \cdot \frac{dL}{dT} \qquad (1)$$

where $α$ [K$^{-1}$] is the true CLTE, $L$ [m] is the linear size of the body, $T$ [K] is the temperature.

It is this parameter that is related to the failure of oxide scale in the processes of deformation and cooling and, accordingly, to the nonuniform conditions on the surface of steel products.

**The goal of this work** is to obtain generalized formulas for calculating the true CLTE of oxide scale in the range from 0 to 1300 ºC, taking into account the movability of critical temperatures in its components.

## METHODS

Research was carried out in three stages: 1) systematization of known data on the thermal expansion for each of the main components of oxide scale at different temperatures; 2) approximation of the systematized data by analytical functions with critical temperatures as formal parameters; 3) convolution of the true CLTE values of individual components into a general formula for scale as a composite material.

**Table 1.** Values of critical temperatures for components of oxide scale at atmospheric pressure

| Component | Essence of critical temperature | Range of variation | References | Basic value* |
|---|---|---|---|---|
| Wüstite | Eutectoid reaction | Equilibrium: 833-873 K (560-600 ºC) | [40-43; 44, p. 15; 45, p. 97] | 843 K (570 ºC) |
| Magnetite | Magnetic transition | 823-900 K (550-627 ºC) | [46, p. 59; 47, p. 254; 48-50] | 848 K (575 ºC) |
| Hematite | Magnetic transition | 943-998 K (670-725 ºC) | [51, p. 5; 52; 53, p. 972; 50; 46, p. 65; 54] | 950 K (677 ºC) |
| Iron | Magnetic transition | 1032-1046 K (759-773 ºC) | [55, p. 46-49; 56, p. 145; 57-59; 61] | 1043 K (770 ºC) |
|  | Polymorphic transformation | Equilibrium: 1183-1208 K (910-935 ºC) | [47, p. 270; 55, p. 46; 56; 60; 61] | 1185 K (912 ºC) |

* The values of critical temperatures that are most common in the technical literature are taken as basic. The formulas in particular form correspond to these basic values (whereas the formulas in general form, also given in the paper, allow taking the different values of critical temperatures in real ranges of their variations).

The most of the data on the thermal expansion for iron oxides are present in the technical literature in the form of the relative elongation of the sample $d$ from the initial length $L^0$ to the current length $L$ at a given temperature $T$:

$$d(T) = \frac{L - L^0}{L^0} \qquad (2)$$

Recalculation of such primary data into the values of the true CLTE $\alpha(T)$ has been performed by the author through preliminary approximation of these data and differentiation of the obtained function, or by the estimated formula:

$$\alpha(T) = \frac{\overline{\alpha_{i+1}}(T_{i+1} - T^0) - \overline{\alpha_i}(T_i - T^0)}{T_{i+1} - T_i} \qquad (3)$$

where $T_i$ and $T_{i+1}$ are tabulated values of temperature in the primary source ($T_i < T_{i+1}$); $T = (T_i + T_{i+1})/2$; $\overline{\alpha_i}$ is the *mean* CLTE between $T^0$ and $T_i$:

$$\overline{\alpha_i} = \frac{d(T_i)}{T_i - T^0} \qquad (4)$$

$\overline{\alpha_{i+1}}$ is the same, between $T^0$ and $T_{i+1}$:

$$\overline{\alpha_{i+1}} = \frac{d(T_{i+1})}{T_{i+1} - T^0} \qquad (5)$$

Two types of phase transitions were taken into account in the second stage of research: polymorphic and magnetic. These types of phase transitions have significant differences due to the fact that during magnetic transitions changes occur only within atoms (electron spins are ordered), but their mutual arrangement in the crystal lattice does not change. Therefore, when choosing approximating functions, the author adopted certain "patterns of behavior" of the true CLTE nearby phase transitions, which are consistent with known empirical data. These "patterns of behavior" assume that at the point of polymorphic transformation the true CLTE experiences a discontinuity (i.e., changes by leaps), while at the point of magnetic transition and at the boundary of thermodynamic stability it changes continuously, but with a possible discontinuity of the first derivative by temperature.

It should be noted that the accepted "patterns of behavior" do not correspond to theoretical provisions: according to the classical theory, the true thermal expansion coefficient should change by a leap at the point of magnetic transition and have an infinite gap at the point of polymorphic transformation [62]. Therefore, these "patterns of behavior" should be considered as reasonable for engineering calculations in conditions of real heterogeneity of composition and structure of materials, as well as blurring of phase transitions in temperature [38; 63].

Taking into account the features of adopted "pattern of behavior" all approximating functions are based on the following general rules:

a) Phase transition temperatures are considered as critical temperatures, which divide the entire target temperature range into separate intervals. For wüstite, its thermodynamic stability boundary (Chaudron point) is considered to be the critical temperature that, according to empirical data, can affect thermal expansion.

b) Within each temperature interval bounded by critical temperatures, the change in properties is described by a smooth function (which derivative does not undergo a discontinuity).

c) At critical temperatures of a magnetic transition or eutectoid reaction (at the Chaudron point for wüstite), the conjugation condition must be satisfied, i.e., the values of the property functions at adjacent intervals must be equal to each other, but the derivatives of these functions can change by leaps (i.e., the property values can change abruptly but continuously).

d) At the critical temperature of the polymorphic transformation, the conjugation condition is not met, i.e., the property values undergo a discontinuity.

e) Within each temperature interval, functions are constructed according to the reference points through which the graph of the function must pass. These are:
- *critical points* and
- *nodal points*.

Nodal points are used to reduce the number of independent parameters that need to be determined by fitting. Points at temperatures 200 K (-73 ºC) и 1600 K (1327 ºC), that are beyond the target temperature range, in most cases, are used as nodal points. The temperature values of the nodal points, as well as the property values at the critical and nodal points are set as independent parameters.

f) Critical temperatures in the formulas are considered as varying parameters. It is assumed that the value of the properties at a given critical point remains constant regardless of the shift in the critical temperature itself. In combination with the unchanged position of the nodal points, this leads to the fact that perturbations due to the shift of the critical temperature are limited only to the region close to it and do not affect the regions far from it.

**RESULTS AND DISCUSSION**

**Magnetite Fe₃O₄**

Known empirical data on the true CLTE of magnetite are shown in **Fig. 1.** These data are approximated separately in two temperature intervals: before and after the magnetic transition point (Curie point). The resulting formulas are summarized in **Table 2** and contain the Curie point ($T_1$) as a formal parameter.

The graphs in **Fig. 2** show the response of these formulas to varying Curie points between 823-900 K (550-627 °C). For example, at 773 K (500 °C), the calculated value of the true CLTE of magnetite is 19.5·10⁻⁶, 18.5·10⁻⁶ or 16.7·10⁻⁶ K⁻¹ for Curie temperature 823, 848 or 900 K respectively. In other words, for different values of the Curie point, the true CLTE of magnetite $\alpha_{Fe3O4}$ at the specified design temperature may differ by 2.8·10⁻⁶ K⁻¹, i.e., by 15% relative to its value at the basic value of 848 K. The results of similar evaluations for different design temperatures in the range of 150 K (±75 K from the Curie basic temperature) are shown in **Table 3**. According to these evaluations, the thermal expansion of magnetite can vary by as much as 30% due to changes in the Curie point over its real range of variation. This example demonstrates the possibility of using critical temperatures (in this case, the Curie point of magnetite) as adaptation parameters in mathematical simulation of processes in the presence of oxide scale on the surface of products.

Fig. 2 also explains the approach to choose the functions passing through independently selected nodal points with coordinates ($T_0$, $y_0$) and ($T_2$, $y_2$) and a critical point with coordinates ($T_1$, $y_1$). The acute-angle connection of the graphs at $T_1$ is chosen considering the accepted "pattern of behavior" of the thermal expansion coefficient at the point of magnetic transition (see above).

At the basic value of the Curie point for magnetite $T_1$ = 848 K (575 ºC), the formulas from Table 2 are reduced to the form ($\alpha_{Fe3O4}$ in [K⁻¹], $T$ in [K]; the corresponding graph is shown in Fig. 1):

– in the range of 273 K ≤ $T$ ≤ 848 K

$$\alpha_{Fe3O4} = \left(-29.571 + 21.181 \cdot T^{0.1} + 10e^{-0.005 \cdot (848-T)}\right) \cdot 10^{-6} \qquad (6)$$

– in the range of 848 K < $T$ ≤ 1573 K

$$\alpha_{Fe3O4} = \left(-20.545 + 1.8564 \cdot T^{0.4} + 15e^{-0.008 \cdot (T-848)}\right) \cdot 10^{-6} \qquad (7)$$

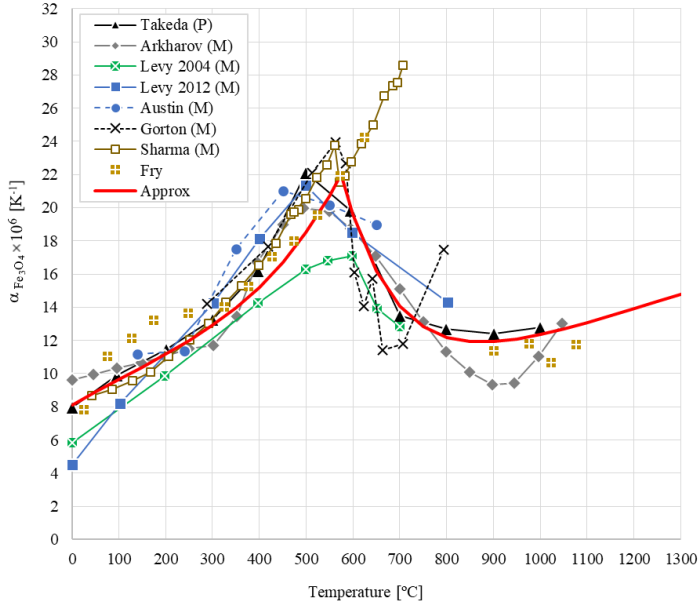

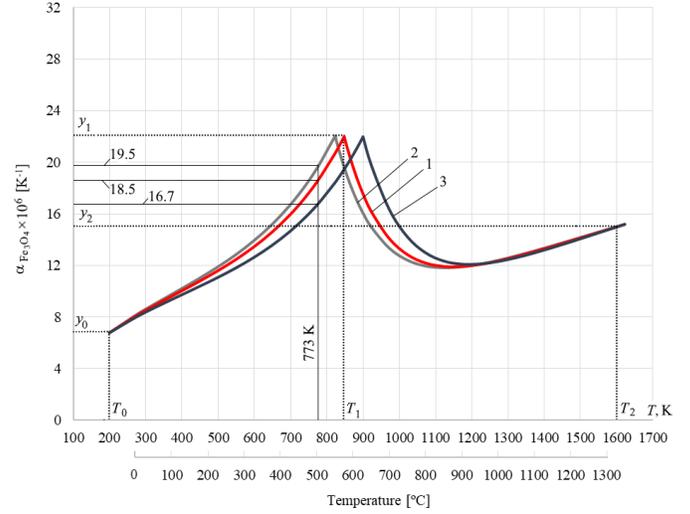

**Fig. 1.** True CLTE of magnetite $\alpha_{Fe3O4}$ according to Takeda[**] [33], Arkharov[**] [64], Levy[**] [65; 66], Austin[*] [67], Gorton[*] [68, p. 279, set 1], Sharma[*] [68, p. 279, sets 2 and 3], Fry[*] [69]. Recalculation of the primary data into the values of the true CLTE was performed by the author of this article (references to such sources are indexed with one asterisk *) or borrowed from [69] (references are indexed with two asterisks **). The conventional symbols in the legend: M - monocrystal, P - polycrystal. Empirical points are conventionally connected by lines. The graph calculated according to (6)-(7) is denoted as "Approx"

**Fig. 2.** Graphs of the true CLTE of $Fe_3O_4$, calculated by the formulas of Table 2 for different values of the Curie point $T_1$: 1 – 848 K (575 °C); 2 – 823 K (550 °C); 3 – 900 K (627 °C).

**Table 3.** Calculated effect of the Curie point of magnetite $T_1$ on its true CLTE

| Design temperature [K] ([°C]) | $\alpha_{Fe3O4} \cdot 10^6$ [K$^{-1}$] at different $T_1$ [K] ([°C]) | | | Range (in % of base - in italics) |
|---|---|---|---|---|
| | 823 (550) | *848 (575)* | 900 (627) | |
| 773 (500) | 19.5 | *18.5* | 16.7 | 15% |
| 848 (575) | 19.6 | *22* | 19.5 | 11% |
| 900 (627) | 16.1 | *17.6* | 22 | 34% |
| 923 (650) | 15.0 | *16.2* | 19.8 | 30% |

**Table 2.** Formulas for calculating the true CLTE of $Fe_3O_4$ ($T_1$ [K] is the Curie point of magnetite)

| Temperature interval [K] | $273 \leq T \leq T_1$ | | |
|---|---|---|---|
| Approximating function [K$^{-1}$] | $\alpha_{Fe3O4}(T) = \left(a_0 + a_1 T^n + a_3 e^{-a_4(T_1-T)}\right) \cdot 10^{-6}$ | | |
| Constants | $n = 0.1$ | $a_3 = 10$ | $a_4 = 0.005$ |
| Coordinates of the reference points | $T_0 = 200$ K | $y_0 = 6.8$ K$^{-1}$ | $y_1 = 22.0$ K$^{-1}$ |
| Coefficients to be calculated | $a_1 = \dfrac{y_1 - y_0 - a_3\left(1 - e^{-a_4(T_1-T_0)}\right)}{T_1^n - T_0^n}$ | | |
| | $a_0 = y_1 - a_1 T_1^n - a_3$ | | |
| Temperature interval [K] | $T_1 < T \leq 1573$ | | |
| Approximating function [K$^{-1}$] | $\alpha_{Fe3O4}(T) = \left(b_0 + b_1 T^p + b_3 e^{-b_4(T-T_1)}\right) \cdot 10^{-6}$ | | |
| Constants | $p = 0.4$ | $b_3 = 15$ | $b_4 = 0.008$ |
| Coordinates of the reference points | $T_2 = 1600$ K | | $y_2 = 15.0$ K$^{-1}$ |
| Coefficients to be calculated | $b_1 = \dfrac{y_1 - y_2 - b_3\left(1 - e^{-b_4(T_2-T_1)}\right)}{T_1^p - T_2^p}$ | | |
| | $b_0 = y_1 - b_1 T_1^p - b_3$ | | |

**Wüstite Fe$_{1-x}$O**

The data presented in the technical publications on the true CLTE of wüstite $\alpha_{FeO}$ (**Fig. 3**) are approximated separately in two temperature intervals: before and after its eutectoid reaction temperature (Chaudron point) $T_1$. The sharp jump in $\alpha_{FeO}$ value between 750 ºC and 550 ºC, recorded by Takeda [33], was not taken into account in the approximation, because it was not observed in other studies.

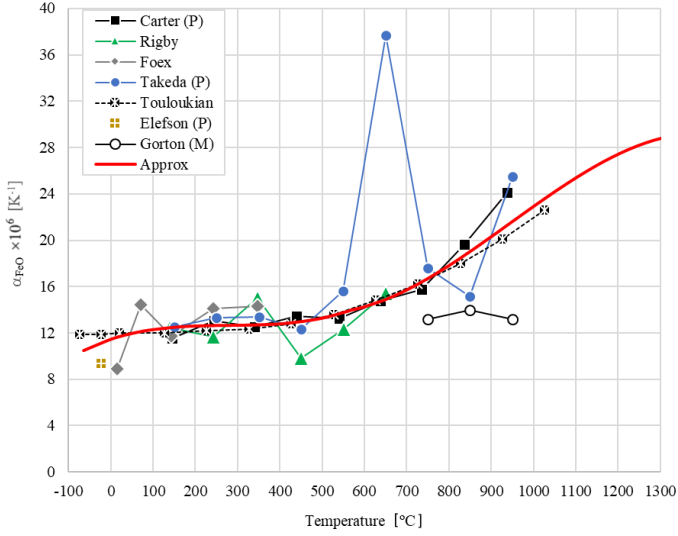

**Fig. 3.** True CLTE of wüstite $\alpha_{FeO}$ according to experimental data of Carter [34], Rigby [71], Foex [72], Takeda [33], Elefson [68, p. 271, set 6], Gorton [68, p. 271, set 1], generalized recommendations of Touloukian [68, p. 271] and calculated by formulas (7)-(8) (denoted as "Approx"). The conventional symbols in the legend: M - monocrystal, P - polycrystal. All empirical values of $\alpha_{FeO}$ were recalculated by the author from primary data of references. The empirical points are conventionally connected by straight lines.

The obtained functions in general form (with the temperature $T_1$ as a formal parameter) are not given in this paper, since the analysis carried out using them shows that shifts in the Chaudron point within real ranges has a very weak effect on the thermal expansion of wüstite. This can be explained by the very gentle course of the graph of the wüstite true CLTE function on the temperature near the Chaudron point (see Fig. 3). Therefore, it makes no practical sense to use the Chaudron point as an adaptation parameter for the thermal expansion coefficient of wüstite. In the particular case, at the basic value $T_1 = 843$ K (570 ºC) the approximating formulas for calculating the true CLTE of wüstite $\alpha_{FeO}$ [K$^{-1}$] as a function of temperature $T$ [K] are as follows (graph - in Fig. 3):

− in the range of 273 K ≤ $T$ ≤ 843 K

$$\alpha_{FeO} = (4.0 + 4.6242 \cdot 10^{-2}T - 8.2889 \cdot 10^{-5}T^2 + 4.9947 \cdot 10^{-8}T^3) \cdot 10^{-6} \quad (8)$$

− in the range of 843 K < $T$ ≤ 1573 K

$$\alpha_{FeO} = (70 - 1.7187 \cdot 10^{-1}T + 1.6258 \cdot 10^{-4}T^2 - 4.4483 \cdot 10^{-8}T^3) \cdot 10^{-6} \quad (9)$$

**Hematite Fe$_2$O$_3$**

Data on the true CLTE of hematite $\alpha_{Fe2O3}$ are shown in **Fig. 4**. Approximating functions in general form (with Curie point $T_1$ as a varying parameter) are given in **Table 4**. According to the generalized formulas presented in this table, the thermal expansion of hematite depends very little on the Curie point shift. For example, Table 5 shows that the difference in $\alpha_{Fe2O3}$ values calculated according to the above formulas for some fixed design temperature at different values of the Curie point $T_1$ does not exceed 5%.

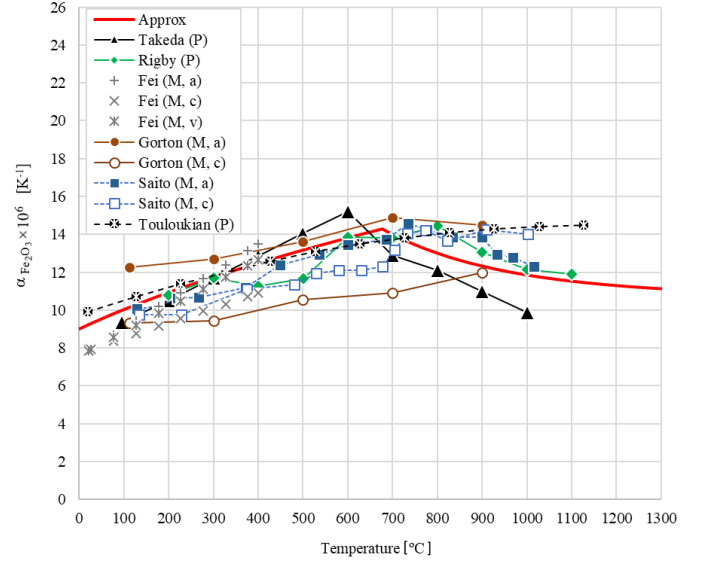

**Fig. 4.** True CLTE of hematite $\alpha_{Fe2O3}$ according to experimental data of Takeda* [33], Rigby* [71], Fei [73, p. 35], Gorton* [68, p. 275], Saito* [68, p. 275], generalized recommendations Touloukian [70, p. 274] and approximating formulas (10)-(11) (denoted as "Approx"). The data in the references with an asterisk * were recalculated into the true coefficient $\alpha_{Fe2O3}$ by the author. The conventional symbols in the legend: P - polycrystal, M - monocrystal, a and c - indices of crystal axes, v - values obtained as 1/3 of the volume expansion coefficient. Empirical points are conventionally connected by lines.

**Table 5.** Calculated effect of the Curie point $T_1$ of hematite on its true CLTE

| Design temperature [K] ([ºC]) | $\alpha_{Fe2O3} \cdot 10^6$ [K$^{-1}$] at different $T_1$ [K] ([ºC]) | | | Range (in % of base - in italics) |
|---|---|---|---|---|
| | 943 (670) | *950 (677)* | 998 (725) | |
| 875 (602) | 13.9 | *13.8* | 13.6 | 2% |
| 950 (677) | 14.3 | *14.3* | 14.0 | 2% |
| 998 (725) | 13.7 | *13.7* | 14.3 | 5% |
| 1025 (752) | 13.4 | *13.4* | 14.0 | 4% |

At the basic value of the Curie point of hematite $T_1 = 950$ K (677 ºC), the approximating formulas for calculating $\alpha_{Fe2O3}$ [K$^{-1}$] as a function of temperature $T$ [K] are as follows:

− in the range of 273 K ≤ $T$ ≤ 950 K

$$\alpha_{Fe2O3} = (2.8760 + 0.37064 \cdot T^{0.5} - 3.2467 \cdot T^{-2}) \cdot 10^{-6} \quad (10)$$

− in the range of 950 K < $T$ ≤ 1573 K

$$\alpha_{Fe2O3} = \left(10.259 + 988.75 \cdot T^{-1} + 3e^{-0.004(T-950)}\right) \cdot 10^{-6} \quad (11)$$

The dependence graph $\alpha_{Fe2O3}(T)$ calculated by (10)-(11) is shown in Fig. 4.

**Table 4.** Formulas for calculating the true CLTE of $Fe_2O_3$ ($T_1$ [K] is the Curie point of hematite)

| Temperature interval [K] | $273 \leq T \leq T_1$ | | |
|---|---|---|---|
| Approximating function [K$^{-1}$] | $\alpha_{Fe2O3}(T) = (a_0 + a_1 T^n + a_2 T^m) \cdot 10^{-6}$ | | |
| Constants | $n = 0.5$ | | $m = -2.0$ |
| Coordinates of the reference points | $T_0 = 273$ K | $y_0 = 9.0$ K$^{-1}$ | $y_1 = 14.3$ K$^{-1}$ |
| Coefficients to be calculated | $a_0 = \dfrac{y_1 T_0^m + y_0 T_1^n - y_0 T_1^m - y_1 T_0^n}{T_0^n T_1^m - T_1^n T_0^m - T_1^m + T_0^m + T_1^n - T_0^n}$ | | |
| | $a_1 = \dfrac{y_0 T_1^m - y_1 T_0^m - a_0(T_1^m - T_0^m)}{T_0^n T_1^m - T_1^n T_0^m}$ | | |
| | $a_2 = \dfrac{y_0 T_1^n - y_1 T_0^n - a_0(T_1^n - T_0^n)}{T_0^m T_1^n - T_1^m T_0^n}$ | | |
| Temperature interval [K] | $T_1 < T \leq 1573$ | | |
| Approximating function [K$^{-1}$] | $\alpha_{Fe2O3}(T) = \left(b_0 + b_1 T^p + b_3 e^{-b_4(T-T_1)}\right) \cdot 10^{-6}$ | | |
| Constants | $p = -1.0$ | $b_3 = 3.0$ | $b_4 = 0.004$ |
| Coordinates of the reference points | $T_2 = 1600$ K | | $y_2 = 11.1$ K$^{-1}$ |
| Coefficients to be calculated | $b_1 = \dfrac{y_1 - y_2 - b_3\left(1 - e^{-b_4(T_2-T_1)}\right)}{T_1^p - T_2^p}$ | | |
| | $b_0 = y_1 - b_1 T_1^p - b_3$ | | |

**Iron Fe**

The known data on the true CLTE of iron $\alpha_{Fe}$ (see **Fig. 5**) are approximated separately in three intervals: below the Curie point $T_1$, between this point and the temperature $T_2$ of polymorphic ($\alpha \leftrightarrow \gamma$) transformation, and above $T_2$. The resulting functions are summarized in **Table 6** and contain temperatures $T_1$ and $T_2$ as formal parameters. These functions are only applicable for $T_1 < T_2$, i.e., under the condition that when the polymorphic transformation shifts (e.g., due to high cooling rate), the magnetic transition still occurs in the alpha phase.

**Tables 7** show the data calculated using the functions from Table 6 to estimate the sensitivity of the CLTE of iron to variations in the Curie point $T_1$. It can be seen that at the same design temperature the thermal expansion of iron can differ by more than 20% depending on the position of the Curie point within its mobility range. The sensitivity of the true CLTE to the variations of the polymorphic transformation temperature is quite obviously determined by the jump of $\alpha_{Fe}$ at point $T_2$ from $16 \cdot 10^{-6}$ to $23 \cdot 10^{-6}$ K$^{-1}$. It should be emphasized that the shift of the temperature of polymorphic transformation $T_2$ for iron in a real technological process with rapid cooling can be much greater than indicated in Table 1 for equilibrium conditions. In other words, it is quite possible to expect a value of $\alpha_{Fe} = 23 \cdot 10^{-6}$ K$^{-1}$ and at temperatures about 800 ºC in the case of high cooling rate.

At the basic values $T_1 = 1043$ K (770 ºC) and $T_2 = 1185$ K (912 ºC), the formulas for the true CLTE of iron $\alpha_{Fe}$ [K$^{-1}$] versus temperature $T$ [K] take the form (graph - in Fig. 5):

− in the range of $273$ K $\leq T \leq 1043$ K

$$\alpha_{Fe} = \left(-21 + 14.765 \cdot T^{0.14} - 7.0642 \cdot e^{-0.013(1043-T)}\right) \cdot 10^{-6} \quad (12)$$

− in the range of $1043$ K $< T \leq 1185$ K

$$\alpha_{Fe} = \left(16.004 - 5.0041 \cdot e^{-0.05(T-1043)}\right) \cdot 10^{-6} \quad (13)$$

− in the range of $1185$ K $< T \leq 1573$ K:

$$\alpha_{Fe} = 23.0 \cdot 10^{-6} \quad (14)$$

**Table 7.** Calculated effect of iron's Curie point $T_1$ on its true CLTE (at polymorphic transformation temperature $T_2 = 1185$ K (912 ºC))

| Design temperature [K] ([ºC]) | $\alpha_{Fe} \cdot 10^6$ [K$^{-1}$] at different $T_1$ [K] ([ºC]) | | | Range (in % of base - in italics) |
|---|---|---|---|---|
| | 1032 (759) | *1043 (770)* | 1046 (773) | |
| 968 (695) | 14.6 | *15.0* | 15.1 | 3% |
| 1032 (759) | 11.0 | *11.9* | 12.1 | 9% |
| 1043 (770) | 13.1 | *11.0* | 11.3 | 19% |
| 1046 (773) | 13.5 | *11.7* | 11.0 | 22% |
| 1118 (845) | 15.9 | *15.9* | 15.9 | 0% |

**Oxide scale as a whole**

**Fig. 6** compares the temperature graphs of the true CLTE for different components of oxide scale, calculated at the basic values of critical temperatures. It can be seen that at the same temperature, these values can differ significantly. Therefore, to estimate the thermal expansion of the entire scale as a whole, it is necessary to "fold" the corresponding values of its components into one formula. Without taking into account porosity in the first approximation this can be done by analogy with Thomas formula for composite materials containing particles of arbitrary shape [76, p. 259]:

$$\alpha_{sc} = [\psi_1 \alpha_1^p + \psi_2 \alpha_2^p + \cdots]^{1/p} \quad (15)$$

where $\alpha_{sc}$ is the true CLTE of oxide scale, $\alpha_1, \alpha_2, \ldots$ are the true CLTE of the 1st, 2nd and other components; $p$ is the exponent; $\psi_1, \psi_2, \ldots$ are the volume fractions of each component without considering pores, i.e.

$$\psi_1 + \psi_2 + \cdots = 1 \quad (16)$$

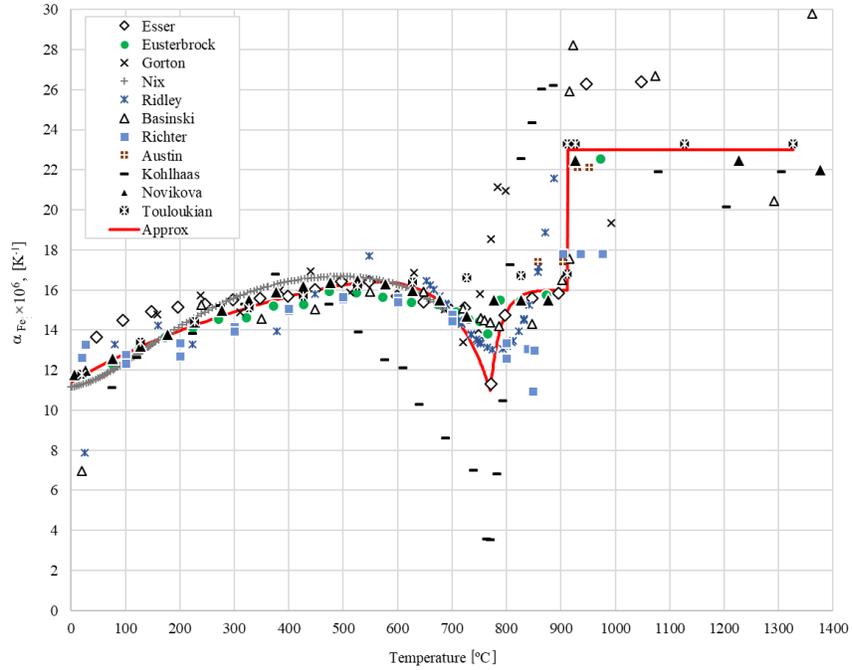

**Fig. 5.** True coefficient of linear thermal expansion of iron $\alpha_{Fe}$ according to the experimental data of Esser [74, p. 198, set 7], Eusterbrock [74, p. 198, set 2], Gorton* [75, p. 158, sets 1 and 2], Nix* [75, p. 158, array 3], Ridley* [75, p. 158, set 16], Basinski* [75, p. 158, set 18], Richter* [75, p. 158, sets 76 and 77], Austin* [75, p. 158, sets 41, 42, 44, 45, and 47], Kohlhaas [74, p. 198, set 5], generalized recommendations from Novikova [74, p. 197] and Touloukian [75, p. 157]. Primary data, references to which are indexed with an asterisk *, were recalculated into values of the true CLTE by the author of this article. The graph calculated by formulas (12)-(14) also is shown (denoted as "Approx")

**Table 6.** Formulas for calculating the true CLTE of iron ($T_1$ [K] is the Curie point of iron; $T_2$ [K] is the point of polymorphic transformation)

| Temperature interval [K] | $273 \leq T \leq T_1$ | | |
|---|---|---|---|
| Approximating function [K$^{-1}$] | $\alpha_{Fe} = \left(a_0 + a_1 T^n + a_3 e^{-a_4(T_1-T)}\right) \cdot 10^{-6}$ | | |
| Constants | $n = 0.14$ | $a_0 = -21.0$ | $a_4 = 0.013$ |
| Coordinates of the reference points | $T_0 = 200$ K | $y_0 = 10.0$ K$^{-1}$ | $y_1 = 11.0$ K$^{-1}$ |
| Coefficients to be calculated | $a_1 = \dfrac{W_1(y_1 - a_0) + a_0 - y_0}{W_1 T_1^n - T_0^n}$ | | |
| | $a_3 = y_1 - a_0 - a_1 T_1^n$ | | |
| Auxiliary parameter | $W_1 = e^{-a_4(T_1 - T_0)}$ | | |
| Temperature interval [K] | $T_1 < T \leq T_2$ | | |
| Approximating function [K$^{-1}$] | $\alpha_{Fe} = \left(b_0 + b_3 e^{-b_4(T-T_1)}\right) \cdot 10^{-6}$ | | |
| Constants | $b_4 = 0.05$ | | |
| Coordinates of the reference points | $y_2 = 16.0$ K$^{-1}$ | | |
| Coefficients to be calculated | $b_3 = \dfrac{y_1 - y_2}{1 - W_2}$ | | |
| | $b_0 = y_1 - b_3$ | | |
| Auxiliary parameter | $W_2 = e^{-b_4(T_2 - T_1)}$ | | |
| Temperature interval [K] | $T_2 < T \leq 1573$ | | |
| Approximating function [K$^{-1}$] | $\alpha_{Fe} = 23.0 \cdot 10^{-6}$ | | |

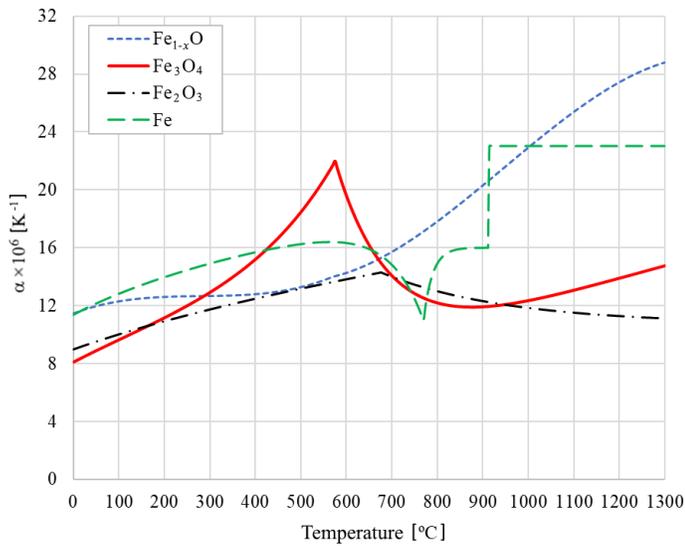

**Fig. 6.** Calculated graphs of the true CLTE of wüstite ($Fe_{1-x}O$), magnetite ($Fe_3O_4$), hematite ($Fe_2O_3$) and iron (Fe) as a function of temperature (calculation was performed according to formulas (8)-(9); (6)-(7); (10)-(11) and (12)-(14) respectively)

The exponent $p$ in formula (15) can take values in the range from -1 to +1, depending on the bulk modulus of elasticity and the mutual arrangement of the components [76, p. 256-260]. Calculations by formula (15) show that if $\alpha_1$, $\alpha_2$, … have the same order of values (as in our case - see Fig. 6), then varying the exponent $p$ from -1 to +1 does not greatly affect the final result. Therefore, in the absence of other data, in the first approximation the $p$ value can be taken close to the middle of the possible range of its variation, i.e., to zero. It can be shown that when $p \to 0$ the function (15) tends to the geometric mean, i.e.:

$$\alpha_{sc} = \alpha_1^{\psi_1} \cdot \alpha_2^{\psi_2} \cdot \ldots \qquad (17)$$

Based on this, the calculation of the true coefficient of linear thermal expansion of oxide scale (excluding pores) is recommended to perform by the formula:

$$\alpha_{sc} = \alpha_{FeO}^{\psi_{FeO}} \cdot \alpha_{Fe3O4}^{\psi_{Fe3O4}} \cdot \alpha_{Fe2O3}^{\psi_{Fe2O3}} \cdot \alpha_{Fe}^{\psi_{Fe}} \cdot \alpha_{xO}^{\psi_{xO}} \qquad (18)$$

where the symbol $\psi$ with an index denotes the volume fraction of the corresponding component in the oxide scale (excluding pores), the symbol $\alpha$ with an index denotes the true CLTE of the corresponding component, and $xO$ stands for the oxide of the alloying element.

As an example, **Fig. 7** shows graphs $\alpha_{sc}$, which are calculated by the formula (18) for the four hypothetical compositions of oxide scale listed in **Table 8**. The first three cases conventionally assume the same component content over the entire temperature range (which can be considered in first approximation as corresponding to the conditions of rapid cooling), and the fourth case has a variable content depending on temperature (modeling the conditions of slow cooling with decomposition of wüstite into an eutectoid mixture of magnetite and metallic iron). In all cases of calculation, the oxides of the alloying elements in the composition of the scale are not taken into account.

**Table 8.** Oxide scale hypothetical compositions for the graphs in Fig. 7.

| Composition | Volume fraction of component $\psi$ | | | |
|---|---|---|---|---|
| | $Fe_{1-x}O$ | $Fe_3O_4$ | $Fe_2O_3$ | Fe |
| 1 | 0,8 | 0,15 | 0,05 | 0 |
| 2 | 0,5 | 0,35 | 0,1 | 0,05 |
| 3 | 0,2 | 0,55 | 0,15 | 0,1 |
| 4 | changes with temperature according to **Fig. 8** | | | |

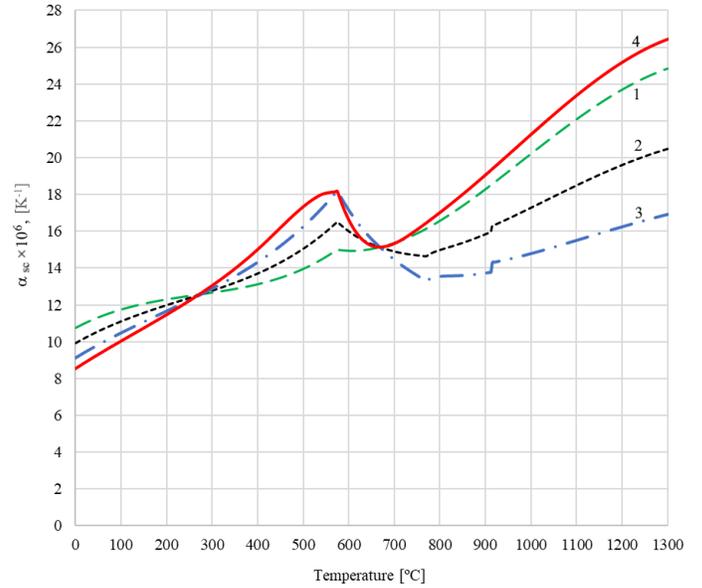

**Fig. 7.** Graphs of the true CLTE of oxide scale calculated by the formula (18) through the corresponding properties of its components (at the basic values of critical temperatures). Numbers in the curves are the numbers of calculated compositions of oxide scale according to Table 8.

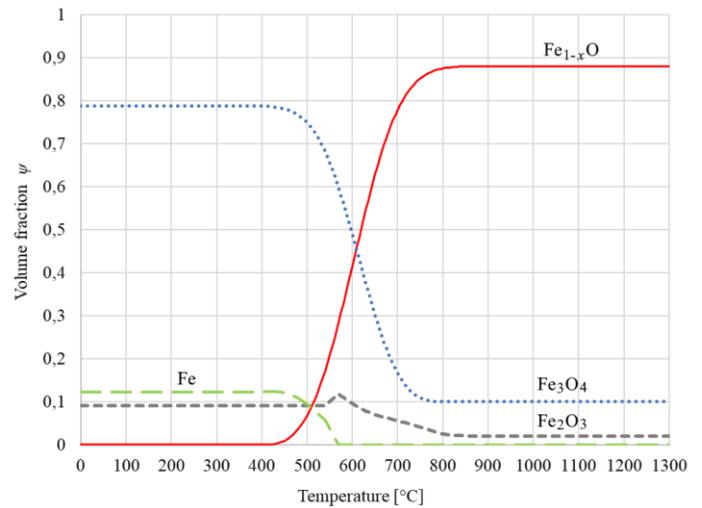

**Fig. 8.** Hypothetical variable oxide scale composition for simulations (number 4 in Fig. 7).

The graphs in **Fig. 7** show that the value of the true CLTE of oxide scale can differ up to 3 times, varying from about $9 \cdot 10^{-6}$ to $26 \cdot 10^{-6}$ K$^{-1}$ in dependence on the temperature and the percentage of different components. It is interesting, that there are two temperatures at which the true CLTE of all main components of oxide scale are close to each other, and the corresponding coefficient for oxide scale is very stable regardless of its percentage composition: at 250 °C $\alpha_{sc} \approx 12 \cdot 10^{-6}$ K$^{-1}$ and at 650 °C $\alpha_{sc} \approx 15 \cdot 10^{-6}$ K$^{-1}$. At the same time, at temperatures from 400 to 600 °C and above 750 °C, the true CLTE of oxide scale can be very unstable, because it depends significantly on the volume fractions of its components.

At the end, let us estimate the influence of the porosity of oxide scale $\eta$, which is the ratio of the pore volume to the total volume of scale, including pores.

In general, the effect of material porosity on thermal expansion is controversial, because it is determined by multidirectional factors. On the one hand, porosity contributes to the reduction of thermal expansion, because the voids can accommodate the "internal" expansion of the material [77]. On the other hand, filling pores with a substance with a higher coefficient of thermal expansion (gases have a coefficient of thermal expansion two orders of magnitude higher than solids) leads to an increase in pressure in the pores as the temperature increases, which contributes to the expansion of the porous body [78-79].

The author is not aware of experimental data on the effect of porosity on the thermal expansion of oxide scale. For other materials, such data are sparse and contradictory. Nevertheless, the results of [72-77] allow us to assume that in the case of interest (crystalline base + gas voids) the first of the above factors prevails, and the dependence of the true linear expansion coefficient of scale on its porosity can be expressed as:

$$\alpha'_{sc} = \alpha_{sc}(1-\eta)^b \qquad (19)$$

where $\boldsymbol{\alpha'_{sc}}$ is "effective" (taking into account pores) value of the true CLTE of oxide scale; $\alpha_{sc}$ - true CLTE of oxide scale without considering pores (by formula (18)); $\eta$ – oxide scale porosity [fractions of one]; $b>0$ is the exponent, in the first approximation it is recommended to take $b \approx 1/3$.

The influence of porosity on the true CLTE of scale by formula (19) at $b=1/3$ is illustrated by the graph in **Fig. 9**. According to experimental data, obtained in the conditions of hot rolling, for furnace scale $\eta \geq 0.2$ and for secondary scale $\eta \approx 0.05$ [2, c. 49-50]. For these values of porosity calculations by the formula (19) show a decrease in the true CLTE by at least 7% for furnace scale and about 2% for secondary one.

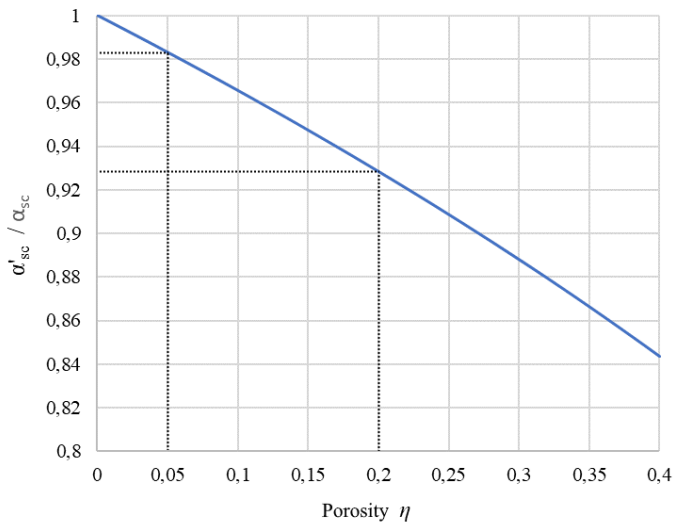

**Fig. 9.** Relative coefficient of influence of scale porosity $\eta$ on the value of its true CLTE (calculation by formula (19) with $b = 1/3$). Dotted lines indicate data corresponding to the level of porosity of furnace and secondary scale during hot rolling.

## CONCLUSION

Abrupt changes in the thermophysical properties of oxide scale components during phase transitions lead to the fact that even a small shift of this critical temperature (due to impurities, defects, particle sizes, etc.) can cause a significant change in the properties of the corresponding component and scale as a whole in a fairly wide temperature range. To account for this factor in engineering calculations, known empirical data on the true coefficient of linear thermal expansion (CLTE) of oxide scale components are generalized and approximated by functions with critical temperatures as formal parameters. The basic assumption of this technique is that the value of the properties at the critical point remains unchanged regardless of its shift in temperature.

By summarizing the empirical data of different authors obtained formulas for calculating the true CLTE in the range from 0 to 1300 °C for the main components of steel scale, namely wüstite ($Fe_{1-x}O$), magnetite ($Fe_3O_4$), hematite ($Fe_2O_3$) and metallic iron. The general formulas obtained allow the possibility to vary the values of critical temperatures and, accordingly, to use them as parameters for adapting mathematical models.

According to the obtained generalized formulas, magnetite has the greatest sensitivity of the thermal expansion to the movability of the Curie point among all the considered components of oxide scale. Thus, at the same design temperature (for example, 650 ºC) the true CLTE of $Fe_3O_4$ can differ by up to 30%, depending on the position of the Curie point within the range of its possible movability. For iron this index is 22%, for hematite - 5%. For wüstite the possible shift of the critical temperature (which is taken as the boundary of its thermodynamic stability - Chaudron point) has almost no effect on its thermal expansion coefficient.

To calculate the true CLTE of oxide scale as a whole ($\alpha_{sc}$), a function of the corresponding values of its components based on Thomas formula for composite materials containing particles of arbitrary shape is proposed. Model calculations show that there are two temperatures at which the true CLTE of all scale components are close to each other, and the corresponding coefficient for oxide scale is very stable regardless of its percentage composition: $\alpha_{sc} \approx 12 \cdot 10^{-6}$ K$^{-1}$ at 250 °C and $\alpha_{sc} \approx 15 \cdot 10^{-6}$ K$^{-1}$ at 650 °C. At the same time, at temperatures from 400 to 600 °C and above 750 °C, the true CLTE of oxide scale can be very unstable, because it depends significantly on the volume fractions of its components.

Based on the analysis of available data on the thermal expansion of various porous materials, a power relationship is proposed to account for the influence of porosity on the thermal expansion of oxide scale.

The proposed methods are recommended for use in the mathematical modeling of the production and processing of steel products in the presence of surface oxide scale.